# Thermodynamic and Relaxation Processes near Curie Point in Gadolinium

Alexander P. Kamantsev[1,a*], Victor V. Koledov[1], Vladimir G. Shavrov[1], Irina S. Tereshina[2]

[1]Kotelnikov Institute of Radio-engineering and Electronics of RAS, Moscow, 125009, Russia
[2]Baikov Institute of Metallurgy and Material Science of RAS, Moscow, 119991, Russia
[a]kama@cplire.ru

**Keywords:** Landau-Khalatnikov equation, Curie point, relaxation time, Gadolinium.

**Abstract.** An experimental method is suggested for the determination of the rate of magnetic phase transitions. The method is based on the measurement of the change of magnetic susceptibility of a ferromagnetic sample in the vicinity of the phase transition in response to an abrupt change of the sample temperature. This paper describes the measurement of the change of the magnetic susceptibility of a thin gadolinium plates, cooled by water-flow at a temperature below the Curie point ($T_C \approx 292$ K). It was found that the relaxation time of the magnetic susceptibility of gadolinium in the temperature range from 289.9 to 291.3 K can be approximated using the Landau-Khalatnikov equation with a kinetic coefficient value $\gamma = 3.9 \times 10^{-8}$ cm$^3$/(erg×s). The linear approximation does not fit well in the range from 291.3 to 293.2 K. The fundamental restriction of specific power of the magnetocaloric refrigerator (made by gadolinium plates) was estimated.

## Introduction

Recently room-temperature magnetic refrigeration is being considered as possible alternative to the conventional Freon refrigerating systems [1]. Magnetic refrigeration is based on magnetic materials with the magnetocaloric effect (MCE). MCE is a phenomenon demonstrated by some magnetic materials, when heating and cooling results from the application and removal of magnetic field. The temperature span of this effect is limited to only few degrees and the largest MCE is observed near a phase transition (PT); for example, this will be about 2 K/T for gadolinium near Curie point temperature [2] and 3.3 K/T for a FeRh alloy near its 1$^{st}$ order PT [3]. The cooling power of magnetic refrigeration is determined by the value of the entropy change in one thermodynamic cycle and the frequency of the thermodynamic cycles. The entropy change of different magnetic compounds is studied in large number of works [4]. However, deep understanding of the relaxation phenomena in PT, which restrict frequency of thermodynamic cycles, does not exist at present. Theoretically, relaxation processes near the 2$^{nd}$ order PT are described by the Landau-Khalatnikov equation [5]:

$$\frac{d\eta}{dt} = -\gamma \frac{\partial \Phi}{\partial \eta} \qquad (1)$$

where $\eta$ – the order parameter, $\Phi$ – the thermodynamic potential per unit volume, $\gamma$ – the kinetic coefficient, $t$ – the time. So far, the validity of Eq. (1) has been proven experimentally for 2$^{nd}$ order PTs such as λ point in liquid helium [5] and nematic-isotropic PT in liquid crystals [6]. However, even for classical magnetic materials near the Curie point, the Landau-Khalatnikov equation has not yet been tested. Thus, the problem of magnetization relaxation near the phase transition in magnetic materials has a significant fundamental and practical importance. The aims of the present work are the following:

1) Development of an experimental method for measuring the relaxation time to the equilibrium value of magnetization following a fast temperature change near PT for magnetocaloric materials.

2) Measurement of magnetization relaxation time near the Curie point in Gd.

3) Estimation of achievable magnitude of specific power of magnetocaloric refrigeration with a working body made of Gd plates.

**Theory**

The general expression for the relaxation time $\tau$ of the order parameter near PT of the 2$^{nd}$ order is given by Lifshitz and Pitaevskii [7]. Following [7] we describe the state of a magnetic body near the Curie point ($T_C$) by the order parameter, i.e. the absolute value of magnetization $M = |\mathbf{M}|$. The magnetization is zero in the paramagnetic state ($T > T_C$), and is non-zero in ferromagnetic state ($T < T_C$). Let us consider the process of relaxation of the magnetization in a system that is not under equilibrium. The equilibrium value of the magnetization $<M>$ is determined by minimizing the appropriate thermodynamic potential. We will use the potential $\Phi$ – a function of temperature $T$ and chemical potential $\mu$ (for a given total volume of the body). The value of $M$ is determined by finding the minimum of $\Phi(T, \mu, M)$ (thermodynamic potential per unit volume) as functions of $M$ at given $T$ and $\mu$ in the spatially homogeneous body:

$$\frac{\partial \Phi}{\partial M} = 0 \tag{2}$$

If this condition is not satisfied, then a process of relaxation occurs, i.e., the parameter $M$ varies with time, approaching $<M>$. In a weakly non-equilibrium state, i.e. at non-zero, but small values of the derivative $\partial\Phi/\partial M$, the rate of relaxation ($\partial M/\partial t$) is also small. In the Landau theory, in which fluctuations of the order parameter are neglected, it must be assumed that the relationship between these two derivatives is reduced to a simple proportionality Eq. (1) with a constant $\gamma$. In Landau theory the thermodynamic potential near the transition point is:

$$\Phi = \Phi_0(T,\mu) + (T - T_C)\alpha M^2 + bM^4 \tag{3}$$

with a positive coefficient $b$; because the ferromagnetic phase corresponds to the region of $T < T_C$, $\alpha > 0$. The equilibrium value of the magnetization in ferromagnetic phase, the solution to Eq. (2) is:

$$<M> = \left[\frac{\alpha(T_C - T)}{2b}\right]^{1/2} \tag{4}$$

The relaxation Eq. (1) becomes:

$$\frac{dM}{dt} = -2\gamma\left[(T - T_C)\alpha M + 2bM^3\right] \tag{5}$$

or by assuming linear behavior of the small difference $\delta M = M - <M>$:

$$\frac{d(\delta M)}{dt} = -\frac{\delta M}{\tau} \tag{6},$$

where

$$\tau = \frac{1}{4\gamma\alpha(T_C - T)}, T < T_C \tag{7}$$

When $t \to \infty$, the difference $\delta M$ should go to zero, and therefore should be $\tau > 0$, and so $\gamma > 0$.

The value of $\tau$ is the magnetization relaxation time. We see, that it goes to infinity when $T \to T_C$. This fact has crucial importance for the whole theory of PT. It ensures the existence of macroscopic states corresponding to incomplete equilibrium for given nonequilibrium values of $M$. It is thanks to this fact, that the magnetization relaxation is considered as independent of the relaxation of other macroscopic characteristics of the body.

The experimental measurement of the relaxation time of a sample is determined practically as the maximum of the thermal exchange time and the magnetic relaxation time. We suggest that the thermal exchange time is equal to the thermal diffusivity time $t_d$ of a sample:

$$t_d \approx \frac{c\rho \times x^2}{\kappa} \tag{8},$$

where $c$ – the heat capacity, $\rho$ – the density, $\kappa$ – the heat conductivity, $x$ – the thickness of a sample in a shape of a plate.

**Experiment**

Gadolinium ingots were prepared by a process of purification by vacuum distillation at optimal regimes. The evaporation of Gd is carried out for 10 h at 420 °K above its melting temperature. The magnetocaloric (MCE) properties of the Gd samples were measured by direct technique in magnetic fields of up to 18 kOe. The results are shown in Fig. 1. Measurement of the MCE show that the maximal adiabatic temperature change $\Delta T_{ad} = 4$ K (in a field of 18 kOe) is observed at 290 K. The Curie point of Gd is near room temperature ($T_C \approx 292$ K), which is very convenient for the experiment. More details were described in [8]. The samples had a shape of a plate with dimensions of 8.5×5.0×0.2 mm. The thermal diffusivity time in Gd plate, calculated from Eq. (8), is $t_d \approx 8$ ms.

The new dynamic thermo-magnetometer (DTM) is proposed for solving the problem of the experimental study the rate of the magnetic PT with response time of less than 10 ms. DTM is designed for measuring the time dependence of the magnetic susceptibility of ferromagnets at an abrupt temperature change in water flow (Fig. 2). The experimental measurements of the magnetic susceptibility of the sample were carried out with the help of a three-coil differential transformer. The number of turns of each measuring coil was 30. An AC signal with a frequency of 2.4 kHz was supplied to the outside coils. The amplitude of the magnetic field excitation is about 1 Oe. The measured signal was taken from the central coil. The selective nanovoltmeter UNIPAN-237 was used to amplify the measured signal. An analog output signal from nanovoltmeter was supplied to a 14-bits analog-to-digital converter L-CARD E14-440, which was connected to the computer. Specialized software allowed to record and to store the results automatically. The temperatures of both samples and water were measured using differential thermocouples with response time of about 10 ms (Fig.2).

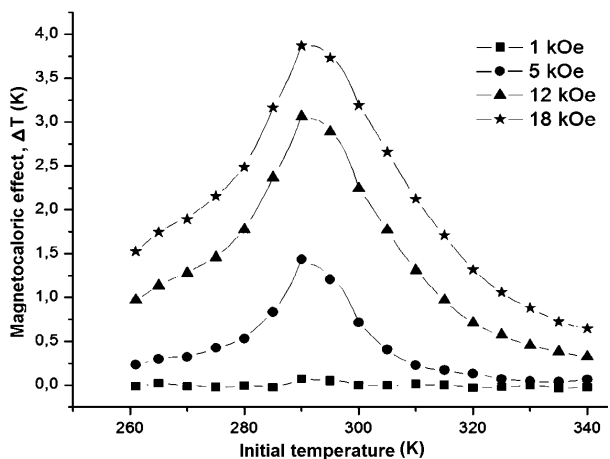
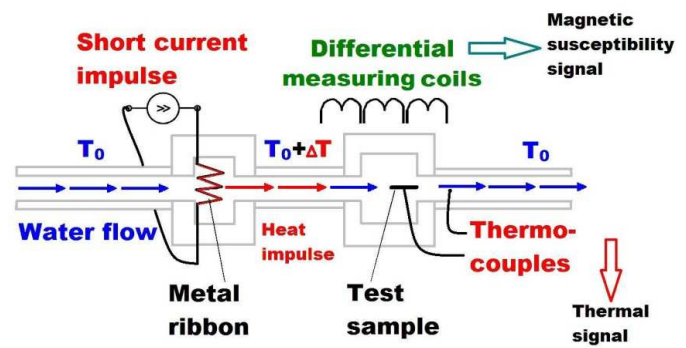

Fig.1. The measured temperature dependence of MCE in different magnetic fields near $T_C$ in Gd.

Fig.2. The scheme of the dynamic thermo-magnetometer.

Gd sample was placed in a chamber under the differential measuring coils. Initially, the water flow was at a temperature of $T_0$, slightly less or slightly more than the Curie temperature of gadolinium. Heat exchange chamber with 35 μm thick metal ribbon was used to heat the water flow abruptly. For this purpose, the metal ribbon was stretched inside the heat exchange chamber, and was heated by short ($\approx$ 15 ms) electrical impulse. The ribbon rapidly transfers heat to the water, forming a "heat impulse". The temperature of the heat impulse was measured by a thermocouple placed in water flow path and had parabolic-like form with a maximum magnitude of 0.4 K. Its duration was about 75 ms (Fig.3). Heat impulse passed Gd sample and heated it. Since, the initial water temperature $T_0$ was lower than the Curie point $T_C$ (ferromagnetic phase), the heat impulse caused a short-time decrease in sample magnetization (Fig. 4), bringing it closer to the Curie point. After that, the heat impulse ends, and the magnetization increases to its original level during magnetic relaxation time. The two typical curves for temperatures $T_0 = 290.5$ K (green) and 291.8 K (yellow) are shown in Fig. 3. The right relaxation part can be exponentially approximated. The relaxation time $\tau$ for each temperature was determined as a fitting parameter of the curve $exp(-t/\tau)$.

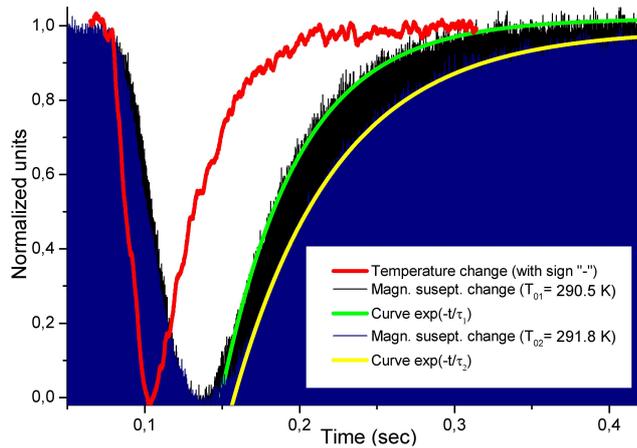 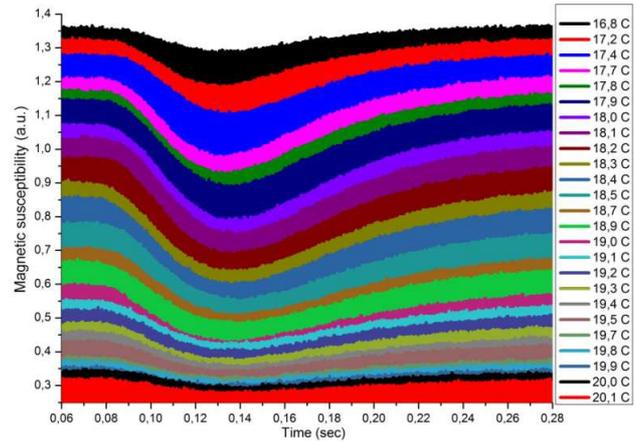

Fig.3. The normalized temporal dependence of water temperature change measured by thermocouple (red) and the comparison of typical curves used to determine the relaxation time of magnetization at different initial temperatures 290.5 K (green), 291.8 K (yellow) of Gd sample.

Fig.4. The temporal changes of magnetic susceptibility of Gd sample at different initial temperatures $T_0$ (from 16.8 to 20.1 C), when heated by a heat impulse.

**Results and discussions**

The results of the series of measurements are presented on Fig. 4. The dependence $1/\tau$ as a function of temperature is shown in Fig.5 after determination of the relaxation time of the magnetization for different initial temperatures of the sample (Fig. 3). The theoretical value of Curie temperature $T_{Cth} = 295.6$ K for Landau-Khalatnikov equation is determined by assuming linear dependence $1/\tau = (const - 4\gamma\alpha T)$ (Fig.5) using Eq. (7). It can be seen that the relaxation time of the magnetic susceptibility of Gd in the temperature range from 289.9 to 291.3 K is well approximated by Landau-Khalatnikov Eq. (7) with a kinetic coefficient value of $\gamma = 3.9\times10^{-8}$ cm$^3$/(erg×s). This value is obtained from Fig. 5 taking for value $\alpha = 2.56\times10^7$ erg/(cm$^3$×K) [10]. The linear assumption is not a good approximation in the temperature range from 291.3 to 293.2 K. We can expect that this is due to critical fluctuations in the vicinity of the Curie point PT. The theoretical value for Curie temperature, $T_{Cth} = 295.6$ K, is higher than the experimental one, $T_C \approx 292$ K, because of these fluctuations existence.

Let us estimate the achievable specific cooling power $P$ of the working body of the magnetic refrigerator working near room temperature. If the working body has a cellular structure consisting of very thin plates [9], then the frequency of cycles is limited to $1/(2\times\tau)$. When T = 290 K the relaxation time is $\tau = 45$ ms (Fig. 5) and the frequency of the cycles of a magnetic refrigerator with a working body made of Gd will be restricted to a value of $f = 1/(2\times\tau) = 11$ Hz. Then taking the entropy change at 2$^{nd}$ order PT in Gd equal to $\Delta S = 5.5$ mJ/(g·K) in magnetic field $H = 20$ kOe [11], we get: $P = \Delta S \times T \times f = 5.5$ mJ/(g×K) × 290 K × 11 Hz ≈ 17.5 W/g.

If we assume that $\tau$ has the same values for the intermetallic compound Fe$_{49}$Rh$_{51}$ with giant reverse MCE, then taking $\Delta S = 11.8$ mJ/(g×K) at T = 297 K in a field of $H = 20$ kOe [3] we get:
$P = \Delta S \times T \times f = 11.8$ mJ/(g×K) × 297 K × 11 Hz ≈ 38.5 W/g.

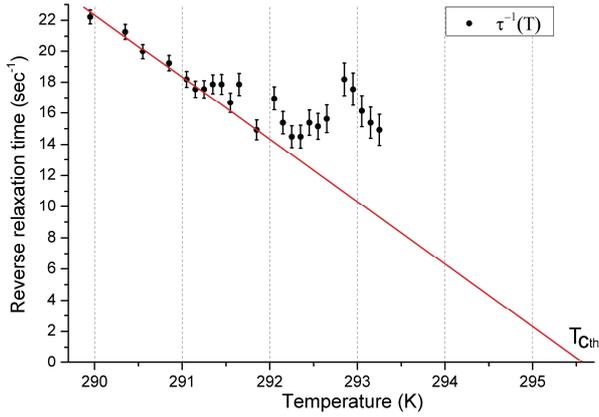 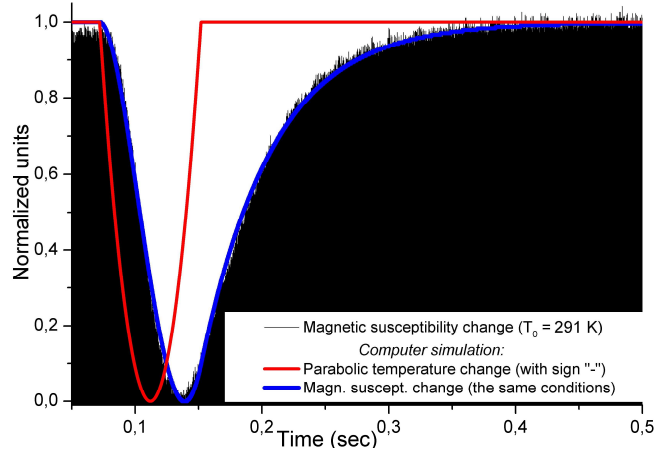

Fig. 5. The dependence of the reverse relaxation time $1/\tau$ on the temperature. The red line is linear approximation: $1/\tau = (const - 4\gamma\alpha T)$. Theoretical Curie point from Landau-Khalatnikov equation is $T_{Cth} = 295.6$ K.

Fig.6. The comparison of experimental data (black) and the computer simulation results of the Landau-Khalatnikov equations with the same conditions in MathCad (blue). Parabolic temperature change with "-"sign is the red curve.

Using Eq. (5) in MathCad, an experiment of the magnetic relaxation was simulated with the temperature depending on time T(t) and with the obtained values of $\gamma$, $T_C = T_{Cth}$. The experimental values of heat impulse (Fig. 3), i.e., the magnitude - 0.4 K, duration - 75 ms and a parabolic-like form were used for of the temperature change simulation. Fig. 6 shows a comparison between the experimental data and the computer simulation results at initial temperature $T_0 = 291$ K. The theoretical and experimental curves are in good agreement at T < 291.3 K (Fig. 6, 7). A discrepancy between the theory and experiment is observed at T > 291.3 K. The theoretical value of $\tau \to \infty$ (Fig. 8) but the experimental value of $\tau$ is approximately constant ~ 60 ms due to critical fluctuations near PT or inhomogeneity of composition of the samples.

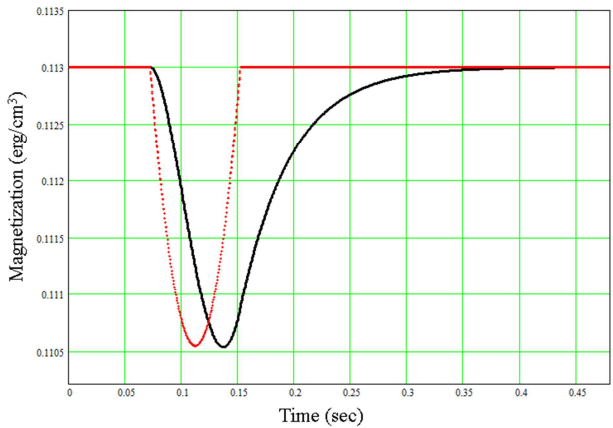 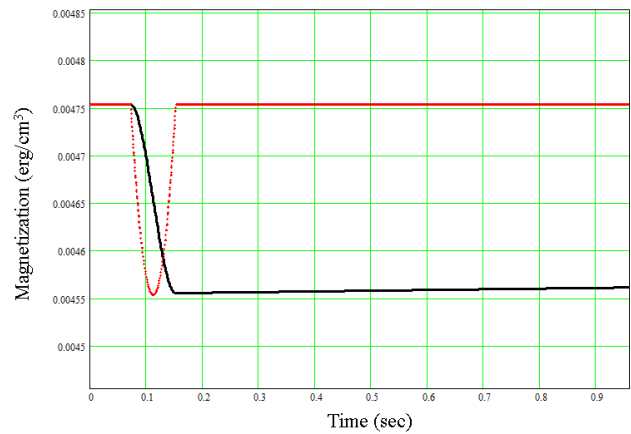

Fig.7. Computer simulation results of magnetization (black) by the Landau-Khalatnikov equations at an initial temperature $T_0 = 290$ K and the normalized parabolic temperature change with "-"sign (red).

Fig.8. Computer simulation results of magnetization (black) by the Landau-Khalatnikov equations at aninitial temperature $T_0 = 295.59$ K and the normalized parabolic temperature change with "-"sign (red).

**Conclusions**

1) The dynamic thermo-magnetometer was constructed for measuring the magnetic susceptibility of ferromagnetic samples versus time near $T_C$ with a time resolution of 10 ms in liquid flow.
2) It is shown that the relaxation time of the magnetic susceptibility of Gd samples in the temperature range of 290 to 291.3 K is well approximated by Landau-Khalatnikov equation with a value $\gamma = 3.9 \times 10^{-8}$ cm$^3$/(erg×s). The linear approximation is not justified in the temperature range of 291.3 to 293.2 K. It is suggested that inhomogeneity of Gd samples or critical fluctuations in the vicinity of the Curie temperature determine the relaxation behavior of the magnetization in this temperature range.

3) The maximum specific power of a magnetocaloric refrigerator with a working body made of thin Gd plates at room temperature, with the frequency of thermodynamic cycles restricted by magnetic relaxation time, is estimated as $P \approx 18$ W/g.

**Acknowledgments**

Authors are thankful to professors P. Ari-Gur and M. Kuz'min for discussions. The reported study was supported by RFBR, research projects No. 12-08-01043 a, 12-08-31340 мол_a.